\documentclass[aps,pra,twocolumn,floatfix]{revtex4}
\usepackage{graphicx}
\usepackage{times}
\usepackage{nicefrac}
\usepackage{amsmath}
\usepackage{amsfonts}
\usepackage{amssymb}
\usepackage{amsthm}
\usepackage{epsf}
\usepackage{bm}
\usepackage{bbm}
\usepackage{color}

\usepackage{dcolumn}
\newcolumntype{.}{D{x}{}{-1}}

\newcommand{\be}{\begin{eqnarray}}
\newcommand{\ee}{\end{eqnarray}}

\newcommand{\eps}{\epsilon}
\newcommand{\veps}{\varepsilon}

\newcommand{\beps}{\bm{\epsilon}}

\newcommand{\bfk}{{\bf k}}

\begin{document}
%
\title{Rayleigh scattering of a linearly polarized light: scenario of the complete experiment}

\author{A. V. Volotka,$^{1,2}$ A. Surzhykov,$^{3,4}$ and S. Fritzsche$^{1,2,5}$}

\affiliation{
$^1$ Helmholtz-Institut Jena, D-07743 Jena, Germany\\
$^2$ GSI Helmholtzzentrum f\"ur Schwerionenforschung, D-64291 Darmstadt, Germany\\
$^3$ Physikalisch-Technische Bundesanstalt, D-38116 Braunschweig, Germany\\
$^4$ Technische Universit\"at Braunschweig, D-38106 Braunschweig, Germany\\
$^5$ Theoretisch-Physikalisches Institut, Friedrich--Schiller-Universit\"at Jena, D-07743 Jena, Germany\\
}

\begin{abstract}
The process of the elastic scattering of photons on atoms, known as the Rayleigh scattering, is investigated. Expressing the scattering observables in terms of the \textit{electric} and \textit{magnetic} complex scattering amplitudes, we work over the scenarios for their independent benchmarking by experiments. In particular, the combination of the cross section and polarization transfer of initially linearly polarized light provides an opportunity for a complete experiment for any but fixed scattering angle. It allows us to deduce the modulus of the electric and magnetic amplitudes together with the phase difference individually. The findings are illustrated by the theoretical calculations of the scattering of $150 - 250$ keV photons on lead atoms.
\end{abstract}


\maketitle

\section{Introduction}
\label{sec:1}

A new generation of experiments has currently become possible due to the recent advent of new solid-state photon detectors. Exploiting the polarization sensitivity of the Compton scattering process, these detectors become highly efficient Compton polarimeters \cite{vetter:2007:363, spillmann:2008:083101, khaplanov:2008:0168, weber:2010:C07010}, which opens a new field to study polarization of x rays. These novel-type position-sensitive x-ray detectors as Compton polarimeters have been exceptionally successfully applied in measuring the linear polarization of photons of the radiative electron capture \cite{tashenov:2006:223202} as well as of the subsequent Ly-$\alpha_1$ emission in the case of capturing in the $L$-shell \cite{weber:2010:243002}. Moreover, they have also been employed in measuring the polarization of the radiation with continuous spectra. In particular, in Ref.~\cite{tashenov:2011:173201}, the linear polarization of bremsstrahlung photons has been measured and, recently, the polarization of the Rayleigh scattered photons has been determined \cite{blumenhagen:2016:119601}.

The Rayleigh scattering being a dominant elastic-scattering process for photon energies less than 2 MeV extensively investigated theoretically, see, e.g., for reviews \cite{kane:1986:75, roy:1999:3, pratt:2005:411}. Relativistic calculations based on the second-order $S$-matrix amplitude is accepted now as a benchmark for treating the Rayleigh scattering \cite{brown:1954:51, johnson:1976:692, kissel:1980:1970, kissel_pratt, roy:1986:1178, manakov:2000:032711, safari:2012:043405, surzhykov:2013:062515, surzhykov:2015:144015}. The angle-differential cross section and the polarization transfer are the typical observables studied theoretically and experimentally. Concerning the polarization studies, for example, in Ref.~\cite{roy:1986:1178}, the polarization of the scattered light for initially unpolarized photon beam was thoroughly compared with experiments. The general properties of the photon-polarization transfer as well as the phenomenon of the circular dichroism in the Rayleigh scattering were addressed in Refs.~\cite{manakov:2000:032711, manakov:2000:4425}. Recently, the numerical studies of the polarization of the scattered photons in case of initially linear polarized light have been made for hydrogen and hydrogenlike ions \cite{safari:2012:043405, surzhykov:2013:062515} as well as for ions with more electrons \cite{surzhykov:2015:144015, omer:2019:113006}.

Despite there are papers devoted to the polarization transfer in the Rayleigh scattering \cite{surzhykov:2015:144015, roy:1986:1178, manakov:2000:4425, manakov:2000:032711, safari:2012:043405, surzhykov:2013:062515, omer:2019:113006} and, in particular, in Refs.~\cite{roy:1986:1178, manakov:2000:032711} significant steps towards complete experiment description were undertaken, we think that the polarization analysis requires more detailed considerations. Further investigations are especially important because of the forthcoming Rayleigh experiment at the PETRA III synchrotron at DESY, where it is planned to go beyond the coplanar geometry employed previously \cite{blumenhagen:2016:119601}. In our present work, we investigate how the polarization properties, together with the cross-section data, could provide an opportunity for the complete description of the scattering process. In particular, we demonstrate how to extract different independent amplitudes, phases from the polarization, and cross-section measurements. For this purpose, in Sec. \ref{sec:2A}, we analyze what kind of parameters can be chosen to describe the scattering process completely. The way of the numerical calculation of the scattering amplitudes is presented in Sec. \ref{sec:2B}. In Sec. \ref{sec:3}, we investigate the angle-differential cross section and polarization of the scattered photons. Special attention is paid to the case of the linearly polarized initial radiation. We find that the polarization analysis and the cross-section data provide a possibility to extract the electric and magnetic amplitudes and the phase difference individually. Numerical calculations are presented for the photons with $150 - 250$ keV energies scattered on lead atoms.

Relativistic units ($\hbar = 1,\,c = 1,\,m = 1$) and the Heaviside charge unit [$\alpha = e^2/(4\pi)$, $e<0$] are used throughout the paper.

\section{Theory}
\label{sec:2}

\subsection{General analysis}
\label{sec:2A}

\begin{figure}
\includegraphics[width=0.45\textwidth]{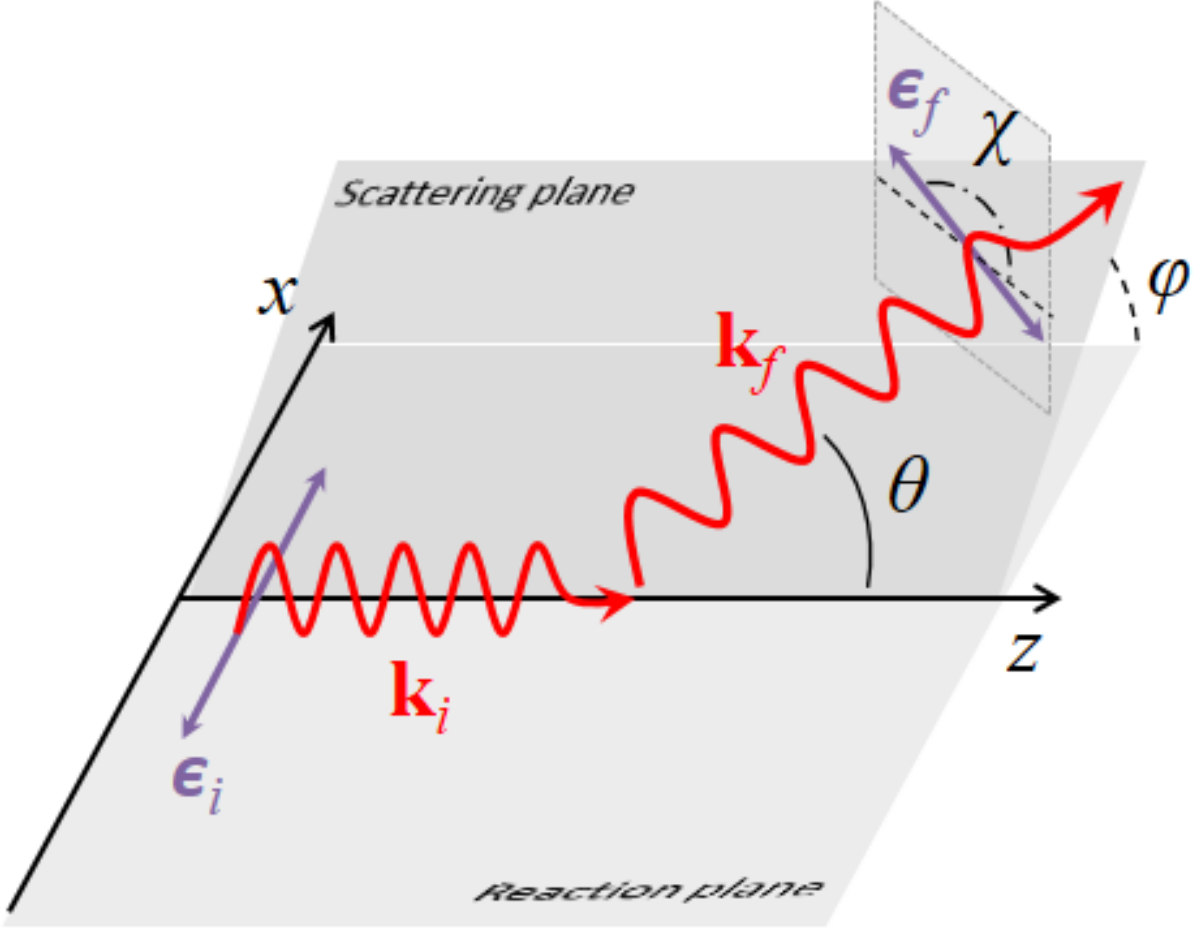}
\caption{(Color online) Geometry of the Rayleigh scattering of a photon. The wave and polarization vectors $\bfk_i$ and $\beps_i$, respectively, of the incoming photons define the reaction plane. The emission direction of the outgoing photons $\bfk_f$ is characterized by the two angles, the scattering $\theta$ and azimuthal $\phi$. The direction of linear polarization of scattered light $\beps_f$ is defined with respect to the scattering plane, as spanned by vectors $\bfk_i$ and $\bfk_f$, and is characterized by the polarization angle $\chi$.
\label{fig:geometry}}
\end{figure}

The general process of the Rayleigh scattering can be illustrated by the following schematic expression:
\begin{equation}
\label{eq:scheme}
A + \gamma_i(\beps_i,\bfk_i) \rightarrow A + \gamma_f(\beps_f,\bfk_f)
\end{equation}
Here, an atom $A$ absorbs the incident photon $\gamma_i$, which is characterized by its momentum $\mathbf{k}_i$ and polarization vector $\boldsymbol{\epsilon}_i$. Due to the photon absorption, an atom undergoes a transition to a virtual state, from which it decays back to its ground state $A$ with the emission of the scattered photon $\gamma_f$ with momentum $\mathbf{k}_f$ and polarization vector $\boldsymbol{\epsilon}_f$. At the same time the energies of the incident and scattered photons are equal, $|\bfk_i| = |\bfk_f| = \omega$. The geometry of the described process is shown in Fig. \ref{fig:geometry}. Two planes can be defined for the process under consideration. These are the scattering plane, the plane spanned by vectors $\mathbf{k}_i$ and $\mathbf{k}_f$, and the reaction plane spanned by vectors $\mathbf{k}_i$ and $\boldsymbol{\epsilon}_i$. The scattering (or polar) angle $\theta$ describes the mutual orientation of the vectors $\mathbf{k}_i$ and $\mathbf{k}_f$ in the scattering plane, while the azimuthal angle $\phi$ is the angle between the reaction and scattering planes. In the case of pure circular initial polarization, the reaction plane is not defined and the dependence on the azimuthal angle $\phi$ disappears.

The amplitude of the process (\ref{eq:scheme}) is the second-order scattering amplitude, which should be linear in both $\beps_i$ and $\beps_f^*$ and include only even number of vectors $\bfk_i$ and $\bfk_f$ \cite{kane:1986:75, manakov:2000:4425}. For atoms with zero total angular momentum, the scattering amplitude ${\cal M}(\beps_i,\bfk_i;\beps_f,\bfk_f)$ can be generally written as:
\be
\label{eq:M_em}
{\cal M}(\beps_i,\bfk_i;\beps_f,\bfk_f) &=& (\beps_i\cdot\beps_f^*) {\cal M}^{({\rm e})}(\theta)\nonumber\\
&+&[\beps_i\times\hat{\bfk}_i]\cdot[\beps_f^*\times\hat{\bfk}_f] {\cal M}^{({\rm m})}(\theta)\,,
\ee
where $\hat{\bfk}$ defines the direction of the photon wave vector $\bfk$. The amplitudes ${\cal M}^{({\rm e})}(\theta)$ and ${\cal M}^{({\rm m})}(\theta)$ are referred to \textit{electric} and \textit{magnetic} scattering amplitudes, respectively. These are two gauge invariant complex functions, which besides the photon energy $\omega$ and atomic structure depend only on the scattering angle $\theta$ or momentum transfer to an atom $q$, $q = 2 \omega \sin{(\theta/2)}$. The reason for the amplitudes names is that in the dipole approximation, the electric amplitude reproduces the electric-dipole polarizability, and the magnetic amplitude reduces to the magnetic-dipole susceptibility \cite{johnson:1968:22}. It can be easily discerned from Eq. (\ref{eq:M_em}).

The electric and magnetic amplitudes can be easily connected with the amplitudes $F(\theta)$ and $G(\theta)$ introduced in previous works \cite{kissel:1980:1970, kane:1986:75, kissel_pratt} by the following expressions:
\be
\label{eq:M_em_FG}
F(\theta) &=& {\cal M}^{({\rm e})}(\theta)
                  + \cos{(\theta)} {\cal M}^{({\rm m})}(\theta)\,,\nonumber\\
G(\theta) &=&-{\cal M}^{({\rm m})}(\theta)\,.
\ee
In addition to the pair of gauge invariant amplitudes introduced above, two other choices are often also employed: linear polarization amplitudes ${\cal M}^\parallel(\theta)$, ${\cal M}^\perp(\theta)$ and circular polarization amplitudes ${\cal M}^{++}(\theta)$, ${\cal M}^{+-}(\theta)$. For the linear amplitudes, polarization vectors $\beps$ are resolved into components parallel $\beps^\parallel$ and perpendicular $\beps^\perp$ to the scattering plane, and the scattering amplitude can be written as
\be
{\cal M}(\beps_i,\bfk_i;\beps_f,\bfk_f) &=& (\beps_i^\parallel\cdot\beps_f^{\parallel *}) {\cal M}^\parallel(\theta)\nonumber\\
&+&(\beps_i^\perp\cdot\beps_f^{\perp *}) {\cal M}^\perp(\theta)\,,
\ee
where the linear amplitudes are given by
\be
\label{eq:M_pp}
{\cal M}^\parallel(\theta) &=& \cos{(\theta)}{\cal M}^{({\rm e})}(\theta)
                            + {\cal M}^{({\rm m})}(\theta)\,,\nonumber\\
{\cal M}^\perp(\theta) &=& {\cal M}^{({\rm e})}(\theta)
                            + \cos{(\theta)}{\cal M}^{({\rm m})}(\theta)\,.
\ee
Another choice is the circular polarization amplitudes, obtained when the polarization vectors $\beps$ are defined as complex circular vector components $\beps^+$ and $\beps^-$,
\be
{\cal M}(\beps_i,\bfk_i;\beps_f,\bfk_f) &=& (\beps_i^+\cdot\beps_f^{+ *} + \beps_i^-\cdot\beps_f^{- *}) {\cal M}^{++}(\theta)\nonumber\\
&+& (\beps_i^+\cdot\beps_f^{- *} + \beps_i^-\cdot\beps_f^{+ *}) {\cal M}^{+-}(\theta)\,,
\ee
where the circular amplitudes can be expressed in terms of the electric and magnetic amplitudes as follows:
\be
\label{eq:M_+-}
{\cal M}^{++}(\theta) &=& \cos^2{\left(\frac\theta2\right)}\left({\cal M}^{({\rm e})}(\theta)
                            + {\cal M}^{({\rm m})}(\theta)\right)\,,\nonumber\\
{\cal M}^{+-}(\theta) &=& \sin^2{\left(\frac\theta2\right)}\left({\cal M}^{({\rm e})}(\theta)
                            - {\cal M}^{({\rm m})}(\theta)\right)\,.
\ee

Thus independently of particular choice of the complex amplitudes there are four real functions, which describe the complete process: ${\cal R}[{\cal M}^{(\rm e)}(\theta)]$, ${\cal I}[{\cal M}^{(\rm e)}(\theta)]$, ${\cal R}[{\cal M}^{(\rm m)}(\theta)]$, and ${\cal I}[{\cal M}^{(\rm m)}(\theta)]$. However, only three of them independent, since the overall phase can not be observed, these are two functions $|{\cal M}^{(\rm e)}(\theta)|$, $|{\cal M}^{(\rm m)}(\theta)|$, and the phase difference $\delta(\theta)$, $\delta(\theta) = {\rm Arg}[{\cal M}^{(\rm e)}(\theta)] - {\rm Arg}[{\cal M}^{(\rm m)}(\theta)]$. Here, we stop our choice on the electric and magnetic amplitudes due to their physical meaning. The electric amplitude describes in the dipole approximation the linear response on the applied electric field and represents the main scattering channel. For its calculation, the form-factor approximation is widely used \cite{franz:1936:314}. In contrast to the electric amplitude, the magnetic amplitude is usually much smaller, and it can not be caught by the form-factor approximation. As a consequence, for its benchmarking by an experiment, a particular scenario should be considered. In what follows, we investigate possibilities for the individual extraction of all scattering functions (the electric and magnetic amplitudes as well as its phase difference) from the scattering observables. However, before we proceed, let us first describe the way we perform the theoretical calculations.

\subsection{$S$-matrix calculation}
\label{sec:2B}

%
\begin{figure*}
\includegraphics[width=\textwidth]{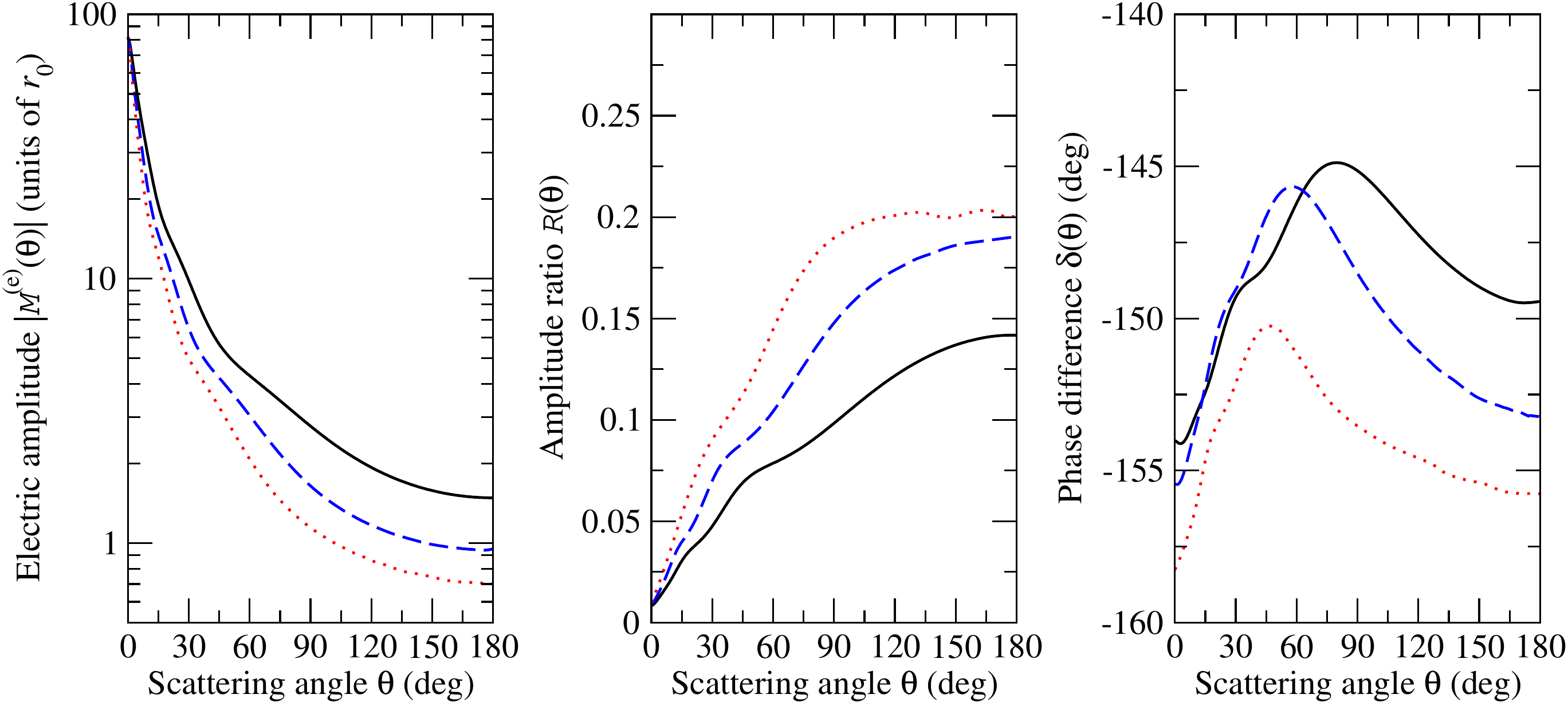}
\caption{(Color online) The electric amplitude (left graph), the amplitude ratio (middle graph), and the phase difference (right graph) as functions of the scattering angle $\theta$. The results of the scattering of photons of $150$ keV (black solid), $200$ keV (blue dashed), and $250$ keV (red dotted) energies on a lead atom are given. The values of electric amplitude are presented in units of the classical electron radius $r_0$.
\label{fig:amplitude}}
\end{figure*}
The scattering amplitude of a whole atom can be found in the independent particle approximation as a sum of scattering amplitude from each of the bound electron. This approximation as was demonstrated in Refs. \cite{lin:1975:1946, volotka:2016:023418} remains rather accurate for a large region of the photon energies. The discussions about the correlation effects can be also found in Refs.~\cite{zhou:1992:6906, fritzsche:2019:1}. Thus, using the $S$-matrix formalism, the scattering amplitude for atoms with closed shells can be written as follows \cite{berestetsky}:
\begin{widetext}
\be
{\cal M}(\beps^{\lambda_i},\bfk_i;\beps^{\lambda_f},\bfk_f) = \dfrac{\alpha}{\omega} \sum \limits_a \sum \limits_n \left( \dfrac{\langle a \vert\boldsymbol{\alpha}\cdot\boldsymbol{\epsilon}^{\lambda_f *}e^{-i\mathbf{k}_f\cdot\mathbf{r}}\vert n\rangle\langle n\vert \boldsymbol{\alpha}\cdot \boldsymbol{\epsilon}^{\lambda_i} e^{i\mathbf{k}_i\cdot\mathbf{r}}\vert a \rangle}{\veps_a+\omega-\veps_n(1-i0)}
+ \dfrac{\langle a \vert\boldsymbol{\alpha}\cdot \boldsymbol{\epsilon}^{\lambda_i} e^{i\mathbf{k}_i\cdot\mathbf{r}}\vert n\rangle\langle n\vert \boldsymbol{\alpha}\cdot \boldsymbol{\epsilon}^{\lambda_f *}e^{-i\mathbf{k}_f\cdot\mathbf{r}}\vert a \rangle}{\veps_a-\omega-\veps_n(1-i0)} \right)\,,
\label{eq:Mfi}
\ee
where $\alpha$ is the fine structure constant, $\boldsymbol{\alpha}$ is the vector of Dirac matrices, $\omega$ stays for the incident and scattered photons energies, $\veps_a$ and $\veps_n$, define the one-electron energies of an active and intermediate state, respectively. The sum over $a$ runs over all occupied (active) one-electron states, and the sum over $n$ runs over the complete set of bound and continuum states of the Dirac equation. The notations $a$ (and $n$) stays for the full set of the relativistic quantum numbers: $n_a$, $\kappa_a$, and $m_a$. The Dirac equation includes not only the Coulomb potential of a nucleus but also a mean-field potential of all atomic electrons. The Eq. (\ref{eq:Mfi}) is rather general, and except the single electron approximation does not relies on any other estimate. However, in the numerical implementation, it is required the complete sum over the one-electron Dirac spectrum. This sum implies the summation over principal quantum numbers as well as over the total angular momentum of the intermediate states. For high-lying atomic shells and high photon energies, the total angular momentum series converge rather poorly. For this reason, in the case of high-lying atomic shells, we employ the following approximation for the amplitude:
\be
{\cal M}(\beps^{\lambda_i},\bfk_i;\beps^{\lambda_f},\bfk_f) = (\beps^{\lambda_i}\cdot\beps^{\lambda_f *}) \sum \limits_a {\cal M}^{{\rm MFF}}_a(\theta)
+ \pi i\dfrac{\alpha}{\omega} \sum \limits_a \sum \limits_{jl} \langle a \vert\boldsymbol{\alpha}\cdot\boldsymbol{\epsilon}^{\lambda_f *}e^{-i\mathbf{k}_f\cdot\mathbf{r}}\vert \eps jl\rangle\langle \eps jl\vert \boldsymbol{\alpha}\cdot \boldsymbol{\epsilon}^{\lambda_i} e^{i\mathbf{k}_i\cdot\mathbf{r}}\vert a \rangle\,,
\label{eq:MFF+}
\ee
\end{widetext}
where
\be
{\cal M}^{{\rm MFF}}_a(\theta) = \int_0^\infty dr r^2 \rho_a(r) \frac{\sin{qr}}{qr}\frac{1}{\veps_a - V_{\rm eff}(r)}
\label{eq:MFF}
\ee
is the modified form-factor amplitude \cite{franz:1936:314, kane:1986:75} and $q$ is the momentum transfer to an atom defined above. The potential $V_{\rm eff}(r)$ is the binding Coulomb potential together with the mean-field electron potential, and $\rho_a(r)$ is the electron charge density of the $a$ electron with the normalization condition $\int_0^\infty dr r^2 \rho_a(r) = 1$. The second term in the equation represents the imaginary part of the scattering amplitude, which corresponds to the photoionization of $a$ electron to a continuum state $\epsilon jl$. The latter characterizes by its energy $\epsilon = \veps_a + \omega$, the total angular momentum $j$, and parity $l$. In contrast to the previously used approximation for high atomic shells \cite{kissel:1980:1970, kane:1986:75, surzhykov:2018:053403, guenther:2018:063843}, we treat the imaginary part of the amplitude exactly.

Numerically most demanding in the computations is the infinite summation over the complete Dirac spectrum $n$, which not only contains the bound states but also the positive- and negative-energy Dirac continuum. In order to perform such a summation, several approaches were employed previously in consideration of the Rayleigh scattering: solution of an inhomogeneous Dirac equation \cite{brown:1954:51}, finite basis-set method \cite{volotka:2016:023418, samoilenko:2020:12}, and exact solution of the Dirac-Coulomb Green’s function \cite{surzhykov:2013:062515, surzhykov:2015:144015, surzhykov:2018:053403}. In the present work, we use the finite basis-set method with the basis constructed from $B$ splines \cite{sapirstein:1996:5213} by employing the dual-kinetic-balance approach \cite{shabaev:2004:130405}.

In Fig. \ref{fig:amplitude} we display the absolute value of electric amplitude $|{\cal M}^{(\rm e)}(\theta)|$, magnetic to electric amplitude ratio
\be
\label{eq:R}
{\cal R}(\theta) = |{\cal M}^{(\rm m)}(\theta)|\,/\,|{\cal M}^{(\rm e)}(\theta)|\,,
\ee
and the phase difference
\be
\label{eq:delta}
\delta(\theta) = {\rm Arg}[{\cal M}^{(\rm e)}(\theta)] - {\rm Arg}[{\cal M}^{(\rm m)}(\theta)]
\ee
as functions of the scattering angle $\theta$. The results obtained for the scattering of $150 - 250$ keV photons on lead atoms. The calculations for the $K$, $L$, $M$, and $N$ shells are performed within the Eq. (\ref{eq:Mfi}), while the scattering on the $O$ and $P$ shells are accounted approximately in terms of Eq. (\ref{eq:MFF+}). These shell restrictions correspond to the corresponding limits in the summation over $a$ in Eqs. (\ref{eq:Mfi}) and (\ref{eq:MFF+}). As the mean-field potential, the self-consistent Kohn-Sham potential \cite{kohn:1965:A1133} is employed. For example, we choose lead atoms, since this target is frequently used for the comparison between theory and experiment, see, e.g., Ref.~\cite{kane:1986:75}. Moreover, the lead atom has zero total angular momentum in its ground state, making Eq.~(\ref{eq:M_em}) exact. Here, however, note, that our approach is well justified for other atoms as well, since most of the atomic electrons are in closed subshells, and the scattering from inner shells is dominant except for small angles where all electrons contribute to scattering \cite{kane:1986:75}. But under small scattering angles, only the electric amplitude describes the process, and extraction of other scattering parameters becomes anyway cumbersome. As one can see from the figure, with an increase of the photon energy, the electric amplitude rapidly decreases with $\theta$, while the amplitude ratio is increasing. The latter agrees with other cases where the magnetic interaction becomes more significant with an increase of the scattering angle for larger energies \cite{goldemberg:1966:311}. We also compared our results with previous calculations \cite{johnson:1976:692, kissel:1980:1970} for the case of 145 keV incident photons scattered on lead atoms and found an agreement within 1-2\%.

\section{Scattering observables}
\label{sec:3}

\subsection{Density matrix of scattered photons}

In order to get access to the scattering observables one has to obtain first the density matrix of the scattered photon $\langle \bfk_f\lambda_f \vert\hat{\rho}_f\vert \bfk_f\lambda_f' \rangle$. With the help of the amplitude derived in the previous section, we can define it as follows \cite{blum,balashov}:
\be
\langle \bfk_f\lambda_f \vert\hat{\rho}_f\vert \bfk_f\lambda_f' \rangle &=& \sum\limits_{\lambda_i\lambda_i'}{\cal M}(\beps^{\lambda_i},\bfk_i;\beps^{\lambda_f},\bfk_f)\langle\lambda_i\vert\hat{\rho}_i\vert\lambda_i'\rangle\nonumber\\
&\times&{\cal M}^*(\beps^{\lambda_i'},\bfk_i;\beps^{\lambda_f'},\bfk_f)\,,
\ee 
where $\langle\lambda_i\vert\hat{\rho}_i\vert\lambda_i'\rangle$ is the density matrix of the initial light. It is convenient to present the density matrix in terms of the Stokes parameters $P_1$, $P_2$, and $P_3$, what yields
\be
\langle \bfk_f\lambda_f \vert\hat{\rho}_f\vert \bfk_f\lambda_f' \rangle = \frac{1}{2}\frac{d\sigma}{d\Omega}\begin{pmatrix}
1 + P_3     & P_1 - i P_2\\
P_1 + i P_2 & 1 - P_3    \\
\end{pmatrix}\,,
\ee
where the prefactor $d\sigma/d\Omega$ is the angle-differential cross section after the summation over the polarization of the scattered light. Here, two Stokes parameters $P_1$ and $P_2$ completely describe the linear polarization of the scattered light. The first parameter is determined by the intensities of light, linearly polarized at a polarization angle $\chi = 0^\circ$ or $\chi = 90^\circ$, $P^{\rm lin}_1 = ( I_{0^\circ} - I_{90^\circ} ) / ( I_{0^\circ} + I_{90^\circ} )$, while a similar ratio but for $\chi = 45^\circ$ or $\chi = 135^\circ$ gives the second parameter $P_2$. Here, the polarization angle $\chi$ is defined with respect to the scattering plane, as spanned by the directions of incident and outgoing photons (cf. Fig. \ref{fig:geometry}). The third Stokes parameter $P_3$ characterizes the degree of circular polarization.

We recall here that our aim is to find scenarios where the individual determination of the scattering parameters, such as electric and magnetic amplitudes as well as the phase difference, becomes possible. Keeping this in mind, we have to fix the polarization of the initial beam.

\subsection{Unpolarized incident light}

%
\begin{figure}
\includegraphics[width=0.45\textwidth]{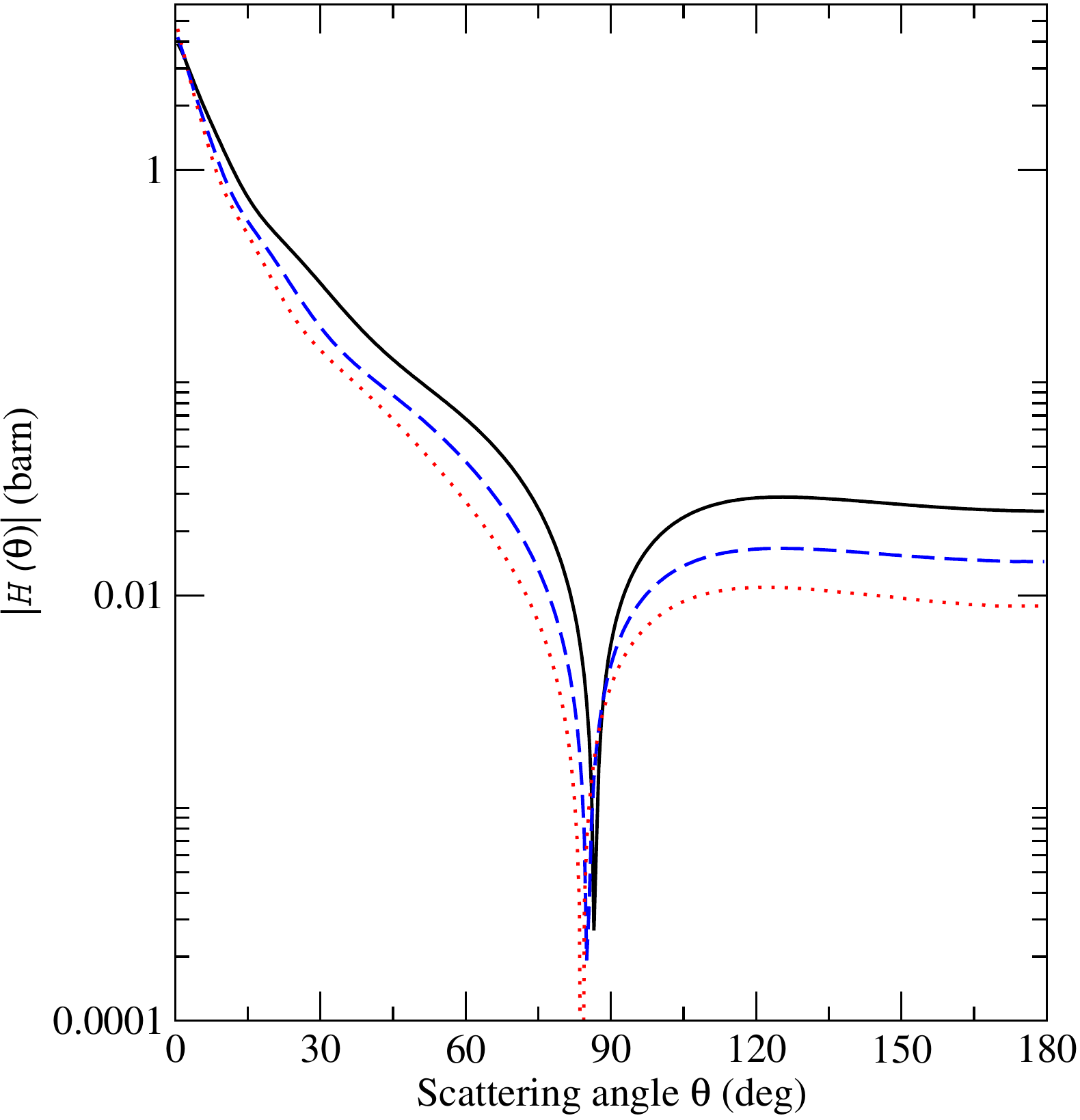}
\caption{(Color online) The modulus of $H(\theta)$ as function of the scattering angle $\theta$, in barn. The scattering of the unpolarized photons of $150$ keV (black solid), $200$ keV (blue dashed), and $250$ keV (red dotted) energies on a lead atoms are considered.
\label{fig:H}}
\end{figure}
Let us start from the case of unpolarized incident radiation. In this case, the angle-differential cross section takes a form:
\be
\frac{d\sigma^{\rm unpol}(\theta,\phi)}{d\Omega} &=& (1+\cos^2{\theta})\left(|{\cal M}^{(\rm e)}(\theta)|^2+|{\cal M}^{(\rm m)}(\theta)|^2\right) \nonumber\\ &+& 4|{\cal M}^{(\rm e)}(\theta)| |{\cal M}^{(\rm m)}(\theta)| \cos{\theta} \cos{\delta}\,,
\ee
and only one nonzero Stokes parameter $P_1$ which characterizes the degree of linear polarization of the scattered photons:
\be
\label{eq:unpol_P1}
P^{\rm unpol}_1(\theta,\phi) = \frac{-\sin^2{\theta}(1-{\cal R}^2)}
           {(1+\cos^2{\theta})(1+{\cal R}^2) + 4{\cal R}\cos{\theta}\cos{\delta}}.
\ee
The polarization $P^{\rm unpol}_1(\theta,\phi)$ given by Eq. (\ref{eq:unpol_P1}) is exactly the polarization transfer $P$ derived previously in Ref.~\cite{roy:1986:1178}. Here, the amplitude ratio ${\cal R}$ and phase difference $\delta$ are defined by Eqs. (\ref{eq:R}) and (\ref{eq:delta}), respectively. Both of them are functions of polar angle $\theta$, which is omitted for the brevity. In the case of zero magnetic amplitude (or in the form-factor approximation) the polarization reduces to simple expression $P^{\rm unpol}_1(\theta,\phi) \approx -\sin^2{\theta} / (1 + \cos^2{\theta})$. Here, we notice that neither cross section $d\sigma^{\rm unpol}/d\Omega$ nor polarization $P^{\rm unpol}_1$ have dependence on the azimuthal angle $\phi$, since for the initial unpolarized radiation the axial symmetry takes place and the reaction plane is not defined anymore. Thus, for each $\theta$ angle, we have only two scattering observables: cross section and one nonzero Stokes parameter. This is definitely not enough to extract all the scattering functions individually. However, if we are interested in the property with the strongest dependence on the magnetic amplitude, we have to consider the following combination $H(\theta)$:
\begin{widetext}
\be
H(\theta) \equiv \displaystyle\frac{d\sigma^{\rm unpol}(\theta,\phi)}{d\Omega} \left( \frac{P^{\rm unpol}_1(\theta,\phi)}{\sin^2{\theta}} + \frac{1}{1+\cos^2{\theta}} \right) =
2|{\cal M}^{(\rm m)}(\theta)|^2 + 4|{\cal M}^{(\rm e)}(\theta)| |{\cal M}^{(\rm m)}(\theta)|
\frac{\cos{\theta} \cos{\delta}}{1+\cos^2{\theta}}\,.
\ee
This function allows effectively to extract information beyond the form factor approximation from the scattering of the unpolarized incident light. The numerical values of this function are presented in Fig. \ref{fig:H}.

\subsection{Linearly polarized incident light}

%
\begin{figure*}
\includegraphics[width=\textwidth]{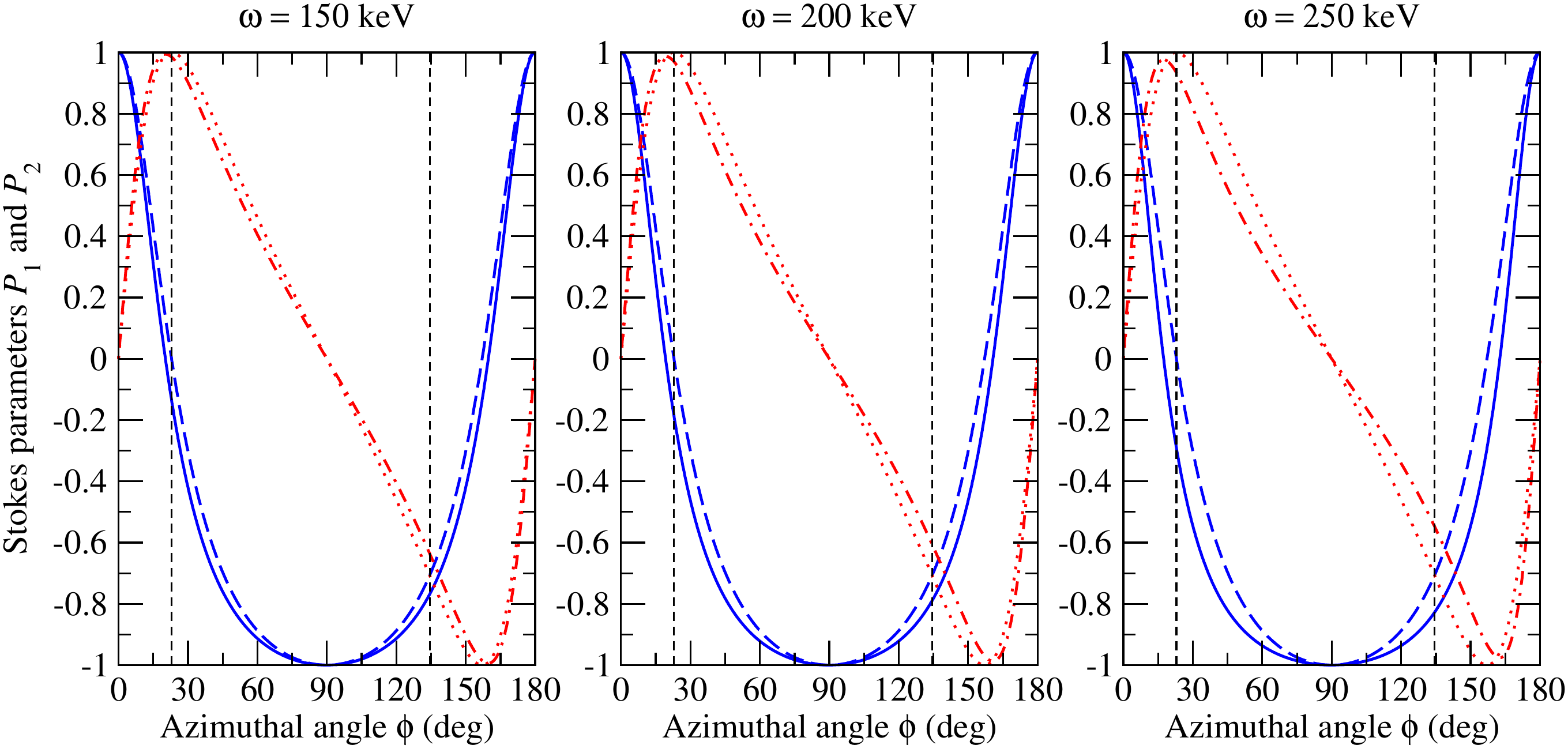}
\caption{(Color online) The Stokes parameters $P_1$ (blue solid and dashed lines) and $P_2$ (red dash-dotted and dotted lines) for the case of initial linear polarization as functions of the azimuthal angle $\phi$ for the fixed scattering angle $\theta = 65^\circ$. The scattering of $150$ keV (left graph), $200$ keV (middle graph), and $250$ keV (right graph) photons on lead atom is evaluated. The approximation when the magnetic amplitude set to zero (blue dashed and red dotted) lines is compared with the calculation where both amplitudes are included (blue solid and red dash-dotted). The dashed vertical lines display azimuthal angles $\phi_z$ and $\phi_c$ suited for the direct determination of the magnetic amplitude (see text for details).
\label{fig:65}}
\end{figure*}
\begin{figure*}
\includegraphics[width=\textwidth]{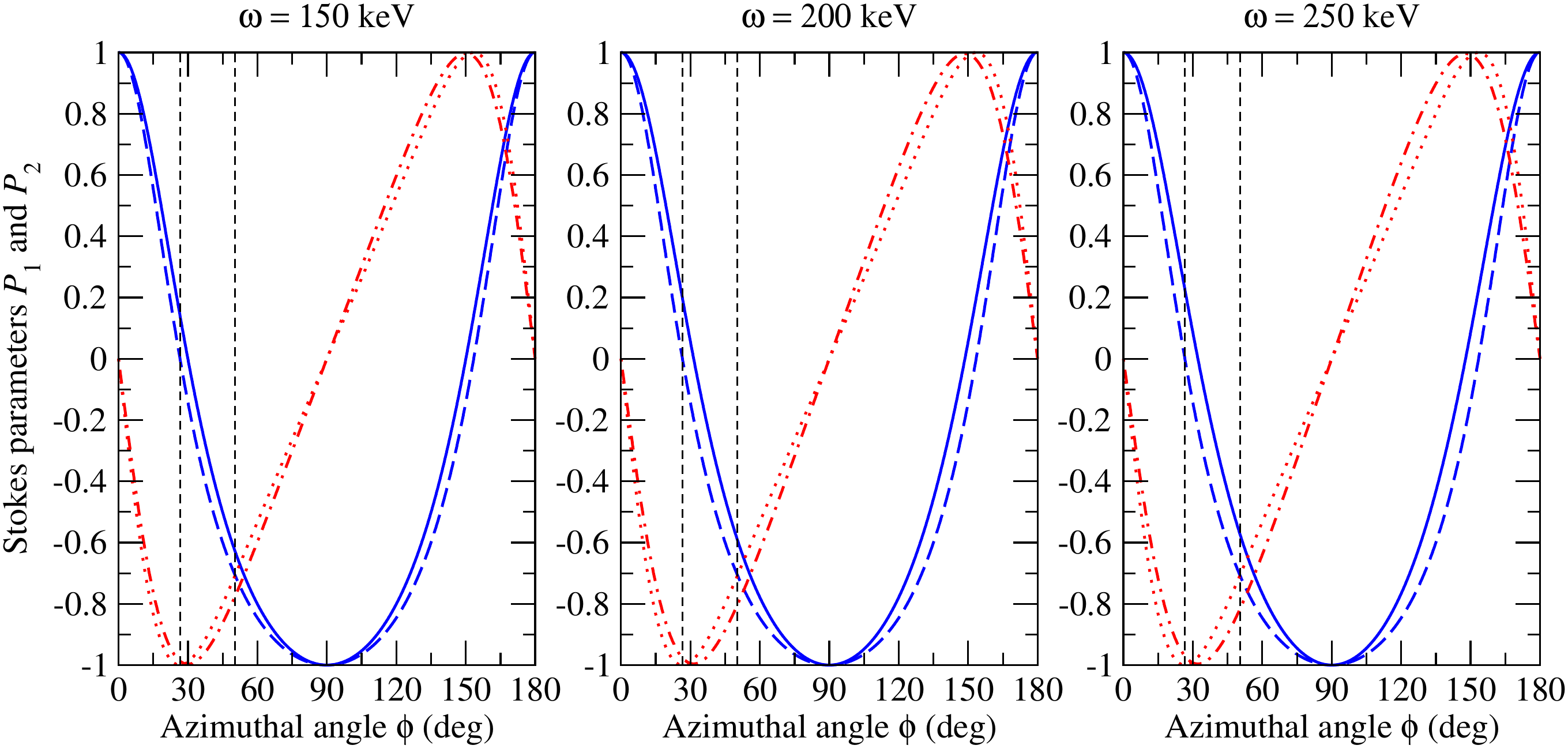}
\caption{(Color online) Same as in Fig. \ref{fig:65}, but for $\theta = 120^\circ$.
\label{fig:120}}
\end{figure*}
Another scenario we would like to consider is when the incident light is completely linearly polarized. This initial polarization is typical for such light sources as synchrotron or free electron laser. Therefore, this scenario requires a special attention. The angle-differential cross section in this case is given by
\be
\label{eq:sig_lin}
\frac{d\sigma^{\rm lin}(\theta,\phi)}{d\Omega} &=& (1+\cos^2{\theta})\left(|{\cal M}^{(\rm e)}(\theta)|^2+|{\cal M}^{(\rm m)}(\theta)|^2\right) - \cos{2\phi}\sin^2{\theta}\left(|{\cal M}^{(\rm e) }(\theta)|^2-|{\cal M}^{(\rm m)}(\theta)|^2\right) \nonumber\\
&+& 4|{\cal M}^{(\rm e)}(\theta)| |{\cal M}^{(\rm m)}(\theta)|\cos{\theta}\cos{\delta}\,,
\ee
while the Stokes parameters take the form:
\be
P^{\rm lin}_1(\theta,\phi) = \frac{\cos{2\phi}(1+\cos^2{\theta})(1+{\cal R}^2) - \sin^2{\theta}(1-{\cal R}^2) + 4{\cal R}\cos{2\phi}\cos{\theta}\cos{\delta}}
  {(1+\cos^2{\theta})(1+{\cal R}^2) - \cos{2\phi}\sin^2{\theta}(1-{\cal R}^2) + 4{\cal R}\cos{\theta}\cos{\delta}}\,,
\ee
\be
P^{\rm lin}_2(\theta,\phi) = \frac{\sin{2\phi}(1+\cos^2{\theta})[\cos{\theta}(1+{\cal R}^2) + (1+\cos^2{\theta}){\cal R}\cos{\delta}]}
                             {(1+\cos^2{\theta})(1+{\cal R}^2) - \cos{2\phi}\sin^2{\theta}(1-{\cal R}^2) + 4{\cal R}\cos{\theta}\cos{\delta}}\,.
\ee
\end{widetext}
We do not present expression for the circular polarization relevant Stokes parameter $P^{\rm lin}_3$, since first the degree of circular polarization much difficult to measure and second it can be deduced from the condition $(P^{\rm lin}_1)^2 + (P^{\rm lin}_2)^2 + (P^{\rm lin}_3)^2 = 1$. Here, we note that the latter is precisely correct only for completely linearly polarized incident light and closed-shell atoms. The dependence of the Stokes on $\phi$ makes them much more rich instruments for the extraction of the scattering parameters. In Figs. \ref{fig:65} and \ref{fig:120} we plot the Stokes parameters as functions of the azimuthal angle $\phi$ for scattering angles $65^\circ$ and $120^\circ$, respectively. In these figures, we compare the results obtained with and without (setting ${\cal R}$ to zero) magnetic amplitude included. As one can see from the figures, there are $\phi$ angles, where the Stokes parameters are especially sensitive to the magnetic amplitude. In particular, for the angle $\phi_z = \arctan{(\pm\cos\theta)}$ the Stokes parameter $P^{\rm lin}_1$ is nonzero only due to the magnetic amplitude: 
\begin{widetext}
\be
P^{\rm lin}_1(\theta,\phi_z) =
  \frac{f_1(\theta,\phi_z){\cal R}^2 + g_1(\theta,\phi_z){\cal R}\cos{\delta}}
       {(1+\cos^2{\theta})(1+{\cal R}^2) - \cos{2\phi_z}\sin^2{\theta}(1-{\cal R}^2) + 4{\cal R}\cos{\theta}\cos{\delta}}\,.
\label{eq:phi_z}
\ee
Another interesting azimuthal angle is $\phi_c = \arctan{[(-1\pm\sqrt{2})\cos{\theta}]}$. In this case, the difference between the second and first Stokes parameters directly depends on the magnetic amplitude:
\be
\Delta P^{\rm lin}(\theta,\phi_c) &\equiv& P^{\rm lin}_2(\theta,\phi_c) - P^{\rm lin}_1(\theta,\phi_c)
\nonumber\\
&=& \frac{f_2(\theta,\phi_c){\cal R}^2 + g_2(\theta,\phi_c){\cal R}\cos{\delta}}
       {(1+\cos^2{\theta})(1+{\cal R}^2) - \cos{2\phi_c}\sin^2{\theta}(1-{\cal R}^2) + 4{\cal R}\cos{\theta}\cos{\delta}}\,.
\label{eq:phi_c}
\ee
Here, functions $f_1$, $f_2$, $g_1$, and $g_2$ are given by
\be
f_1(\theta,\phi) = \cos{2\phi}(1+\cos^2{\theta}) + \sin^2{\theta}\,,
\ee
\be
g_1(\theta,\phi) = 4\cos{2\phi}\cos{\theta}\,,
\ee
\be
f_2(\theta,\phi) = (\sin{2\phi}\cos{\theta} - \cos{2\phi})(1+\cos^2{\theta}) - \sin^2{\theta}\,,
\ee
\be
g_2(\theta,\phi) = 4[\sin{2\phi}(1+\cos^2{\theta})-\cos{2\phi}\cos{\theta}]\,.
\ee
These angles $\phi_z$ and $\phi_c$ we represent by dashed vertical lines in Figs. \ref{fig:65} and \ref{fig:120}. Combing now Eqs. (\ref{eq:phi_z}) and (\ref{eq:phi_c}) with the corresponding result for the cross section (\ref{eq:sig_lin}) one can extract the magnetic amplitude:
\be
|{\cal M}^{(\rm m)}(\theta)|^2 =
\frac{
  g_2(\theta,\phi_c)\displaystyle\frac{d\sigma^{\rm lin}(\theta,\phi_z)}{d\Omega}P^{\rm lin}_1(\theta,\phi_z)
- g_1(\theta,\phi_z)\displaystyle\frac{d\sigma^{\rm lin}(\theta,\phi_c)}{d\Omega}\Delta P^{\rm lin}(\theta,\phi_c)
     }
{ g_2(\theta,\phi_c) f_1(\theta,\phi_z) - g_1(\theta,\phi_z) f_2(\theta,\phi_c) }\,.
\ee
Knowing now the magnetic amplitude we can extract the electric one as follows:
\be
|{\cal M}^{(\rm e)}(\theta)|^2 =
  \frac{d\sigma^{\rm lin}(\theta,\phi)}{d\Omega}
  \frac{\cos{2\phi} - P^{\rm lin}_1(\theta,\phi)}{(1-\cos{2\phi})\sin^2{\theta}}
+ |{\cal M}^{(\rm m)}(\theta)|^2\,,
\ee
here, $\phi$ is not restricted to any particular angle. Finally, we can also determine the phase:
\be
\cos{\delta} =
\frac{1}{|{\cal M}^{(\rm e)}(\theta)| |{\cal M}^{(\rm m)}(\theta)|}
\frac{
  f_2(\theta,\phi_c)\displaystyle\frac{d\sigma^{\rm lin}(\theta,\phi_z)}{d\Omega}P^{\rm lin}_1(\theta,\phi_z)
- f_1(\theta,\phi_z)\displaystyle\frac{d\sigma^{\rm lin}(\theta,\phi_c)}{d\Omega}\Delta P^{\rm lin}(\theta,\phi_c)
     }
     { f_2(\theta,\phi_c) g_1(\theta,\phi_z) - f_1(\theta,\phi_z) g_2(\theta,\phi_c) }\,.
\ee
\end{widetext}
Here, one should say that we determine only the cosine of the phase difference. Thus, one can not extract its sign. The complementary dependence on the $\sin\delta$ can only be identified by considering the incident (partial) circularly polarized photons or measuring the circular polarization of the scattered light, i.e., $P_3$ parameter. However, for hundreds of keV photon energies, there are not many sources of circularly polarized photons available, and a measurement of the circular degree of polarization is somewhat tricky. One of the solutions could be to repeat the scattering but use the scattered photons as the incident. In this case, the $P_2$ parameter of the second time scattered photons depends on the $\sin\delta$ and can be used to determine the sign of the phase difference.

\begin{figure}
\includegraphics[width=0.5\textwidth]{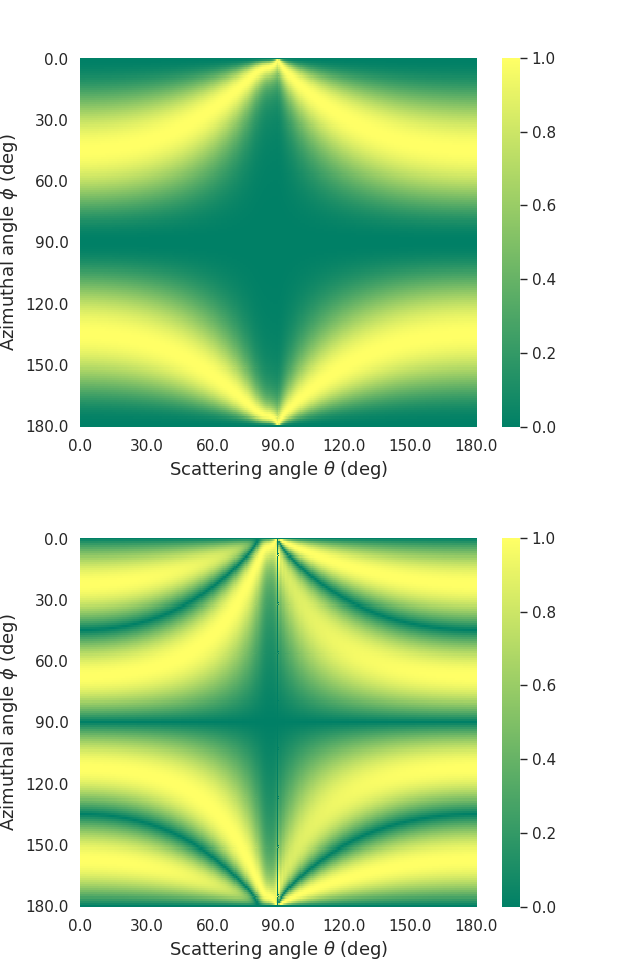}
\caption{(Color online) Sensitivity of the Stokes parameters ($P^{\rm lin}_1$ - top chart, $P^{\rm lin}_2$ - bottom chart) to the magnetic amplitude as functions of azimuthal $\phi$ and scattering $\theta$ angles. The sensitivity, being the normalized difference between the Stokes parameters calculated with and without magnetic amplitude, varies from 0 (small sensitivity) to 1 (large sensitivity). The color bars are depicted right from the charts. The data for the scattering of 200 keV photons on a lead atom is used.
\label{fig:P1P2}}
\end{figure}

Thus, we demonstrate that the scattering parameters such as the electric and magnetic amplitudes and the phase difference can be extracted from the $\phi$-dependence analysis of the scattered lights' polarization properties. In principle, one can employ for the scattering parameters determination also other $\phi$ angles. However, one has to keep in mind that the magnetic amplitude is much smaller than its electric counterpart, so special care should be taken. In other words, we have to find $\phi$ angles with maximum sensitivity to the magnetic amplitude. As one can see from Figs. \ref{fig:65} and \ref{fig:120} the $\phi$ angles with maximum sensitivity depend on the scattering angle $\theta$. Therefore, in Fig. \ref{fig:P1P2} we investigate the sensitivity of the Stokes parameters $P^{\rm lin}_1$ and $P^{\rm lin}_2$ on the ratio ${\cal R}$ for different scattering and azimuthal angles. The sensitivity is defined as the normalized difference between Stokes parameters evaluated with and without (setting ${\cal R}$ to zero) magnetic amplitude included. As one can see from Fig. \ref{fig:P1P2}, for the first Stokes parameter the most sensitive angles $\phi = 45^\circ$ and $135^\circ$ for forward and backward scattering changes to $\phi = 0^\circ$ and $180^\circ$ for $\theta = 90^\circ$. For the second Stokes parameter there are four angles for forward and backward scattering, $\phi = 22.5^\circ$, $67.5^\circ$, $112.5^\circ$, and $157.5^\circ$, which reduce to $\phi = 0^\circ$ and $180^\circ$ for $\theta = 90^\circ$.

We do not consider the case of initial circular polarization since there are no instruments that could provide such a light beam with hundreds of keV photon energy.

\section{Summary}
\label{sec:4}

Summarizing, we have theoretically investigated the Rayleigh scattering for the closed-shell atoms. The different sets of the independent amplitudes which could completely describe the Rayleigh scattering process have been discussed. Absolute values of the electric and magnetic amplitudes, as well as the phase difference between them, have been chosen as the scattering parameter set for a complete description of the process. We have studied the polarization properties of the scattered light aiming for the extraction of these parameters independently. We have shown that the polarization transfer of the initially linear polarized light and the corresponding cross-section data provide a comprehensive tool for determining the scattering parameters. The azimuthal angles, where the dependence on the magnetic amplitude is maximized, have been proposed, and the sensitivity of other azimuthal angles have been investigated numerically. The findings obtained will be beneficial for the Rayleigh scattering experiments, which can be performed with the synchrotron radiation sources, where the radiation is almost entirely linearly polarized \cite{blumenhagen:2016:119601}.

Moreover, the detailed knowledge of the scattering parameters is needed in other respects as well. In particular, in Refs.~\cite{tashenov:2009:599, surzhykov:2018:053403}, it was shown that the Rayleigh scattering could be used as a susceptible tool for the x-ray polarimetry. In order to extract the polarization of the initial beam, the scattering process has to be understood in many details. Another issue is the possibility for the observation of the Delbr\"uck scattering for 1 MeV photons proposed in Ref. \cite{koga:2017:204801}. Here, as in the case of polarimetry, an important ingredient of the successful experiment is the detailed knowledge of the Rayleigh scattering parameters.

\section*{Acknowledgments}

We thank Wilko Middents and G\"unter Weber for valuable discussions.



\begin{thebibliography}{40}
\expandafter\ifx\csname natexlab\endcsname\relax\def\natexlab#1{#1}\fi
\expandafter\ifx\csname bibnamefont\endcsname\relax
  \def\bibnamefont#1{#1}\fi
\expandafter\ifx\csname bibfnamefont\endcsname\relax
  \def\bibfnamefont#1{#1}\fi
\expandafter\ifx\csname citenamefont\endcsname\relax
  \def\citenamefont#1{#1}\fi
\expandafter\ifx\csname url\endcsname\relax
  \def\url#1{\texttt{#1}}\fi
\expandafter\ifx\csname urlprefix\endcsname\relax\def\urlprefix{URL }\fi
\providecommand{\bibinfo}[2]{#2}
\providecommand{\eprint}[2][]{\url{#2}}

\bibitem[{\citenamefont{Vetter}(2007)}]{vetter:2007:363}
\bibinfo{author}{\bibfnamefont{K.}~\bibnamefont{Vetter}},
  \bibinfo{journal}{Annu. Rev. Nucl. Part. Sci.} \textbf{\bibinfo{volume}{57}},
  \bibinfo{pages}{363} (\bibinfo{year}{2007}).

\bibitem[{\citenamefont{Spillmann et~al.}(2008)\citenamefont{Spillmann,
  Br\"auning, Hess, Beyer, St\"ohlker, Dousse, Protic, and
  Krings}}]{spillmann:2008:083101}
\bibinfo{author}{\bibfnamefont{U.}~\bibnamefont{Spillmann}},
  \bibinfo{author}{\bibfnamefont{H.}~\bibnamefont{Br\"auning}},
  \bibinfo{author}{\bibfnamefont{S.}~\bibnamefont{Hess}},
  \bibinfo{author}{\bibfnamefont{H.}~\bibnamefont{Beyer}},
  \bibinfo{author}{\bibfnamefont{T.}~\bibnamefont{St\"ohlker}},
  \bibinfo{author}{\bibfnamefont{J.-C.} \bibnamefont{Dousse}},
  \bibinfo{author}{\bibfnamefont{D.}~\bibnamefont{Protic}}, \bibnamefont{and}
  \bibinfo{author}{\bibfnamefont{T.}~\bibnamefont{Krings}},
  \bibinfo{journal}{Rev. Sci. Instrum.} \textbf{\bibinfo{volume}{79}},
  \bibinfo{pages}{083101} (\bibinfo{year}{2008}).

\bibitem[{\citenamefont{Khaplanov et~al.}(2008)\citenamefont{Khaplanov,
  Tashenov, Cederwall, and Jaworski}}]{khaplanov:2008:0168}
\bibinfo{author}{\bibfnamefont{A.}~\bibnamefont{Khaplanov}},
  \bibinfo{author}{\bibfnamefont{S.}~\bibnamefont{Tashenov}},
  \bibinfo{author}{\bibfnamefont{B.}~\bibnamefont{Cederwall}},
  \bibnamefont{and} \bibinfo{author}{\bibfnamefont{G.}~\bibnamefont{Jaworski}},
  \bibinfo{journal}{Nucl. Instr. and Meth. A} \textbf{\bibinfo{volume}{593}},
  \bibinfo{pages}{459} (\bibinfo{year}{2008}).

\bibitem[{\citenamefont{Weber et~al.}(2010{\natexlab{a}})\citenamefont{Weber,
  Br\"auning, Hess, M\"artin, Spillmann, and St\"ohlker}}]{weber:2010:C07010}
\bibinfo{author}{\bibfnamefont{G.}~\bibnamefont{Weber}},
  \bibinfo{author}{\bibfnamefont{H.}~\bibnamefont{Br\"auning}},
  \bibinfo{author}{\bibfnamefont{S.}~\bibnamefont{Hess}},
  \bibinfo{author}{\bibfnamefont{R.}~\bibnamefont{M\"artin}},
  \bibinfo{author}{\bibfnamefont{U.}~\bibnamefont{Spillmann}},
  \bibnamefont{and}
  \bibinfo{author}{\bibfnamefont{T.}~\bibnamefont{St\"ohlker}},
  \bibinfo{journal}{J. Instrum.} \textbf{\bibinfo{volume}{5}},
  \bibinfo{pages}{C07010} (\bibinfo{year}{2010}{\natexlab{a}}).

\bibitem[{\citenamefont{Tashenov et~al.}(2006)\citenamefont{Tashenov,
  St\"ohlker, Bana\'s, Beckert, Beller, Beyer, Bosch, Fritzsche, Gumberidze,
  Hagmann et~al.}}]{tashenov:2006:223202}
\bibinfo{author}{\bibfnamefont{S.}~\bibnamefont{Tashenov}},
  \bibinfo{author}{\bibfnamefont{T.}~\bibnamefont{St\"ohlker}},
  \bibinfo{author}{\bibfnamefont{D.}~\bibnamefont{Bana\'s}},
  \bibinfo{author}{\bibfnamefont{K.}~\bibnamefont{Beckert}},
  \bibinfo{author}{\bibfnamefont{P.}~\bibnamefont{Beller}},
  \bibinfo{author}{\bibfnamefont{H.~F.} \bibnamefont{Beyer}},
  \bibinfo{author}{\bibfnamefont{F.}~\bibnamefont{Bosch}},
  \bibinfo{author}{\bibfnamefont{S.}~\bibnamefont{Fritzsche}},
  \bibinfo{author}{\bibfnamefont{A.}~\bibnamefont{Gumberidze}},
  \bibinfo{author}{\bibfnamefont{S.}~\bibnamefont{Hagmann}},
  \bibnamefont{et~al.}, \bibinfo{journal}{Phys. Rev. Lett.}
  \textbf{\bibinfo{volume}{97}}, \bibinfo{pages}{223202}
  (\bibinfo{year}{2006}).

\bibitem[{\citenamefont{Weber et~al.}(2010{\natexlab{b}})\citenamefont{Weber,
  Br\"auning, Surzhykov, Brandau, Fritzsche, Geyer, Hagmann, Hess, Kozhuharov,
  M\"artin et~al.}}]{weber:2010:243002}
\bibinfo{author}{\bibfnamefont{G.}~\bibnamefont{Weber}},
  \bibinfo{author}{\bibfnamefont{H.}~\bibnamefont{Br\"auning}},
  \bibinfo{author}{\bibfnamefont{A.}~\bibnamefont{Surzhykov}},
  \bibinfo{author}{\bibfnamefont{C.}~\bibnamefont{Brandau}},
  \bibinfo{author}{\bibfnamefont{S.}~\bibnamefont{Fritzsche}},
  \bibinfo{author}{\bibfnamefont{S.}~\bibnamefont{Geyer}},
  \bibinfo{author}{\bibfnamefont{S.}~\bibnamefont{Hagmann}},
  \bibinfo{author}{\bibfnamefont{S.}~\bibnamefont{Hess}},
  \bibinfo{author}{\bibfnamefont{C.}~\bibnamefont{Kozhuharov}},
  \bibinfo{author}{\bibfnamefont{R.}~\bibnamefont{M\"artin}},
  \bibnamefont{et~al.}, \bibinfo{journal}{Phys. Rev. Lett.}
  \textbf{\bibinfo{volume}{105}}, \bibinfo{pages}{243002}
  (\bibinfo{year}{2010}{\natexlab{b}}).

\bibitem[{\citenamefont{Tashenov et~al.}(2011)\citenamefont{Tashenov, B\"ack,
  Barday, Cederwall, Enders, Khaplanov, Poltoratska, Sch\"assburger, and
  Surzhykov}}]{tashenov:2011:173201}
\bibinfo{author}{\bibfnamefont{S.}~\bibnamefont{Tashenov}},
  \bibinfo{author}{\bibfnamefont{T.}~\bibnamefont{B\"ack}},
  \bibinfo{author}{\bibfnamefont{R.}~\bibnamefont{Barday}},
  \bibinfo{author}{\bibfnamefont{B.}~\bibnamefont{Cederwall}},
  \bibinfo{author}{\bibfnamefont{J.}~\bibnamefont{Enders}},
  \bibinfo{author}{\bibfnamefont{A.}~\bibnamefont{Khaplanov}},
  \bibinfo{author}{\bibfnamefont{Y.}~\bibnamefont{Poltoratska}},
  \bibinfo{author}{\bibfnamefont{K.-U.} \bibnamefont{Sch\"assburger}},
  \bibnamefont{and}
  \bibinfo{author}{\bibfnamefont{A.}~\bibnamefont{Surzhykov}},
  \bibinfo{journal}{Phys. Rev. Lett.} \textbf{\bibinfo{volume}{107}},
  \bibinfo{pages}{173201} (\bibinfo{year}{2011}).

\bibitem[{\citenamefont{Blumenhagen et~al.}(2016)\citenamefont{Blumenhagen,
  Fritzsche, Gassner, Gumberidze, M\"artin, Schell, Seipt, Spillmann,
  Surzhykov, Trotsenko et~al.}}]{blumenhagen:2016:119601}
\bibinfo{author}{\bibfnamefont{K.-H.} \bibnamefont{Blumenhagen}},
  \bibinfo{author}{\bibfnamefont{S.}~\bibnamefont{Fritzsche}},
  \bibinfo{author}{\bibfnamefont{T.}~\bibnamefont{Gassner}},
  \bibinfo{author}{\bibfnamefont{A.}~\bibnamefont{Gumberidze}},
  \bibinfo{author}{\bibfnamefont{R.}~\bibnamefont{M\"artin}},
  \bibinfo{author}{\bibfnamefont{N.}~\bibnamefont{Schell}},
  \bibinfo{author}{\bibfnamefont{D.}~\bibnamefont{Seipt}},
  \bibinfo{author}{\bibfnamefont{U.}~\bibnamefont{Spillmann}},
  \bibinfo{author}{\bibfnamefont{A.}~\bibnamefont{Surzhykov}},
  \bibinfo{author}{\bibfnamefont{S.}~\bibnamefont{Trotsenko}},
  \bibnamefont{et~al.}, \bibinfo{journal}{New J. Phys.}
  \textbf{\bibinfo{volume}{18}}, \bibinfo{pages}{119601}
  (\bibinfo{year}{2016}).

\bibitem[{\citenamefont{Kane et~al.}(1986)\citenamefont{Kane, Kissel, Pratt,
  and Roy}}]{kane:1986:75}
\bibinfo{author}{\bibfnamefont{P.~P.} \bibnamefont{Kane}},
  \bibinfo{author}{\bibfnamefont{L.}~\bibnamefont{Kissel}},
  \bibinfo{author}{\bibfnamefont{R.~H.} \bibnamefont{Pratt}}, \bibnamefont{and}
  \bibinfo{author}{\bibfnamefont{S.~C.} \bibnamefont{Roy}},
  \bibinfo{journal}{Phys. Rep.} \textbf{\bibinfo{volume}{140}},
  \bibinfo{pages}{75} (\bibinfo{year}{1986}).

\bibitem[{\citenamefont{Roy et~al.}(1999)\citenamefont{Roy, Kissel, and
  Pratt}}]{roy:1999:3}
\bibinfo{author}{\bibfnamefont{S.~C.} \bibnamefont{Roy}},
  \bibinfo{author}{\bibfnamefont{L.}~\bibnamefont{Kissel}}, \bibnamefont{and}
  \bibinfo{author}{\bibfnamefont{R.~H.} \bibnamefont{Pratt}},
  \bibinfo{journal}{Rad. Phys. Chem.} \textbf{\bibinfo{volume}{56}},
  \bibinfo{pages}{3} (\bibinfo{year}{1999}).

\bibitem[{\citenamefont{Pratt}(2005)}]{pratt:2005:411}
\bibinfo{author}{\bibfnamefont{R.~H.} \bibnamefont{Pratt}},
  \bibinfo{journal}{Rad. Phys. Chem.} \textbf{\bibinfo{volume}{74}},
  \bibinfo{pages}{411} (\bibinfo{year}{2005}).

\bibitem[{\citenamefont{Brown et~al.}(1954)\citenamefont{Brown, Peierls, and
  Woodward}}]{brown:1954:51}
\bibinfo{author}{\bibfnamefont{G.~E.} \bibnamefont{Brown}},
  \bibinfo{author}{\bibfnamefont{R.~E.} \bibnamefont{Peierls}},
  \bibnamefont{and} \bibinfo{author}{\bibfnamefont{J.~B.}
  \bibnamefont{Woodward}}, \bibinfo{journal}{Proc. R. Soc. Lond. A}
  \textbf{\bibinfo{volume}{227}}, \bibinfo{pages}{51} (\bibinfo{year}{1954}).

\bibitem[{\citenamefont{Johnson and Cheng}(1976)}]{johnson:1976:692}
\bibinfo{author}{\bibfnamefont{W.~R.} \bibnamefont{Johnson}} \bibnamefont{and}
  \bibinfo{author}{\bibfnamefont{K.-T.} \bibnamefont{Cheng}},
  \bibinfo{journal}{Phys. Rev. A} \textbf{\bibinfo{volume}{13}},
  \bibinfo{pages}{692} (\bibinfo{year}{1976}).

\bibitem[{\citenamefont{Kissel et~al.}(1980)\citenamefont{Kissel, Pratt, and
  Roy}}]{kissel:1980:1970}
\bibinfo{author}{\bibfnamefont{L.}~\bibnamefont{Kissel}},
  \bibinfo{author}{\bibfnamefont{R.~H.} \bibnamefont{Pratt}}, \bibnamefont{and}
  \bibinfo{author}{\bibfnamefont{S.~C.} \bibnamefont{Roy}},
  \bibinfo{journal}{Phys. Rev. A} \textbf{\bibinfo{volume}{22}},
  \bibinfo{pages}{1970} (\bibinfo{year}{1980}).

\bibitem[{\citenamefont{Kissel and Pratt}(1985)}]{kissel_pratt}
\bibinfo{author}{\bibfnamefont{L.}~\bibnamefont{Kissel}} \bibnamefont{and}
  \bibinfo{author}{\bibfnamefont{R.~H.} \bibnamefont{Pratt}},
  \emph{\bibinfo{title}{Rayleigh Scattering. Elastic Photon Scattering by Bound
  Electrons}} (\bibinfo{publisher}{{{\rm in} Atomic Inner-shell Physics, {\rm
  Ed. by B.~Crasemann, pp. 465 - 498}, Plenum Press, New York}},
  \bibinfo{year}{1985}).

\bibitem[{\citenamefont{Roy et~al.}(1986)\citenamefont{Roy, Sarkar, Kissel, and
  Pratt}}]{roy:1986:1178}
\bibinfo{author}{\bibfnamefont{S.~C.} \bibnamefont{Roy}},
  \bibinfo{author}{\bibfnamefont{B.}~\bibnamefont{Sarkar}},
  \bibinfo{author}{\bibfnamefont{L.~D.} \bibnamefont{Kissel}},
  \bibnamefont{and} \bibinfo{author}{\bibfnamefont{R.~H.} \bibnamefont{Pratt}},
  \bibinfo{journal}{Phys. Rev. A} \textbf{\bibinfo{volume}{34}},
  \bibinfo{pages}{1178} (\bibinfo{year}{1986}).

\bibitem[{\citenamefont{Manakov
  et~al.}(2000{\natexlab{a}})\citenamefont{Manakov, Meremianin, Carney, and
  Pratt}}]{manakov:2000:032711}
\bibinfo{author}{\bibfnamefont{N.~L.} \bibnamefont{Manakov}},
  \bibinfo{author}{\bibfnamefont{A.~V.} \bibnamefont{Meremianin}},
  \bibinfo{author}{\bibfnamefont{J.~P.~J.} \bibnamefont{Carney}},
  \bibnamefont{and} \bibinfo{author}{\bibfnamefont{R.~H.} \bibnamefont{Pratt}},
  \bibinfo{journal}{Phys. Rev. A} \textbf{\bibinfo{volume}{61}},
  \bibinfo{pages}{032711} (\bibinfo{year}{2000}{\natexlab{a}}).

\bibitem[{\citenamefont{Safari et~al.}(2012)\citenamefont{Safari, Amaro,
  Fritzsche, Santos, Tashenov, and Fratini}}]{safari:2012:043405}
\bibinfo{author}{\bibfnamefont{L.}~\bibnamefont{Safari}},
  \bibinfo{author}{\bibfnamefont{P.}~\bibnamefont{Amaro}},
  \bibinfo{author}{\bibfnamefont{S.}~\bibnamefont{Fritzsche}},
  \bibinfo{author}{\bibfnamefont{J.~P.} \bibnamefont{Santos}},
  \bibinfo{author}{\bibfnamefont{S.}~\bibnamefont{Tashenov}}, \bibnamefont{and}
  \bibinfo{author}{\bibfnamefont{F.}~\bibnamefont{Fratini}},
  \bibinfo{journal}{Phys. Rev. A} \textbf{\bibinfo{volume}{86}},
  \bibinfo{pages}{043405} (\bibinfo{year}{2012}).

\bibitem[{\citenamefont{Surzhykov et~al.}(2013)\citenamefont{Surzhykov,
  Yerokhin, Jahrsetz, Amaro, {Th.~St\"ohlker}, and
  Fritzsche}}]{surzhykov:2013:062515}
\bibinfo{author}{\bibfnamefont{A.}~\bibnamefont{Surzhykov}},
  \bibinfo{author}{\bibfnamefont{V.~A.} \bibnamefont{Yerokhin}},
  \bibinfo{author}{\bibfnamefont{T.}~\bibnamefont{Jahrsetz}},
  \bibinfo{author}{\bibfnamefont{P.}~\bibnamefont{Amaro}},
  \bibinfo{author}{\bibnamefont{{Th.~St\"ohlker}}}, \bibnamefont{and}
  \bibinfo{author}{\bibfnamefont{S.}~\bibnamefont{Fritzsche}},
  \bibinfo{journal}{Phys. Rev. A} \textbf{\bibinfo{volume}{88}},
  \bibinfo{pages}{062515} (\bibinfo{year}{2013}).

\bibitem[{\citenamefont{Surzhykov et~al.}(2015)\citenamefont{Surzhykov,
  Yerokhin, {Th.~St\"ohlker}, and Fritzsche}}]{surzhykov:2015:144015}
\bibinfo{author}{\bibfnamefont{A.}~\bibnamefont{Surzhykov}},
  \bibinfo{author}{\bibfnamefont{V.~A.} \bibnamefont{Yerokhin}},
  \bibinfo{author}{\bibnamefont{{Th.~St\"ohlker}}}, \bibnamefont{and}
  \bibinfo{author}{\bibfnamefont{S.}~\bibnamefont{Fritzsche}},
  \bibinfo{journal}{J. Phys. B} \textbf{\bibinfo{volume}{48}},
  \bibinfo{pages}{144015} (\bibinfo{year}{2015}).

\bibitem[{\citenamefont{Manakov
  et~al.}(2000{\natexlab{b}})\citenamefont{Manakov, Meremianin, Maquet, and
  Carney}}]{manakov:2000:4425}
\bibinfo{author}{\bibfnamefont{N.~L.} \bibnamefont{Manakov}},
  \bibinfo{author}{\bibfnamefont{A.~V.} \bibnamefont{Meremianin}},
  \bibinfo{author}{\bibfnamefont{A.}~\bibnamefont{Maquet}}, \bibnamefont{and}
  \bibinfo{author}{\bibfnamefont{J.~P.~J.} \bibnamefont{Carney}},
  \bibinfo{journal}{J. Phys. B} \textbf{\bibinfo{volume}{33}},
  \bibinfo{pages}{4425} (\bibinfo{year}{2000}{\natexlab{b}}).

\bibitem[{\citenamefont{Omer and Hajima}(2019)}]{omer:2019:113006}
\bibinfo{author}{\bibfnamefont{M.}~\bibnamefont{Omer}} \bibnamefont{and}
  \bibinfo{author}{\bibfnamefont{R.}~\bibnamefont{Hajima}},
  \bibinfo{journal}{New J. Phys.} \textbf{\bibinfo{volume}{21}},
  \bibinfo{pages}{113006} (\bibinfo{year}{2019}).

\bibitem[{\citenamefont{Johnson and Feiock}(1968)}]{johnson:1968:22}
\bibinfo{author}{\bibfnamefont{W.~R.} \bibnamefont{Johnson}} \bibnamefont{and}
  \bibinfo{author}{\bibfnamefont{F.~D.} \bibnamefont{Feiock}},
  \bibinfo{journal}{Phys. Rev.} \textbf{\bibinfo{volume}{168}},
  \bibinfo{pages}{22} (\bibinfo{year}{1968}).

\bibitem[{\citenamefont{Franz}(1936)}]{franz:1936:314}
\bibinfo{author}{\bibfnamefont{W.}~\bibnamefont{Franz}}, \bibinfo{journal}{Z.
  Phys.} \textbf{\bibinfo{volume}{98}}, \bibinfo{pages}{314}
  (\bibinfo{year}{1936}).

\bibitem[{\citenamefont{Lin et~al.}(1975)\citenamefont{Lin, Cheng, and
  Johnson}}]{lin:1975:1946}
\bibinfo{author}{\bibfnamefont{C.-P.} \bibnamefont{Lin}},
  \bibinfo{author}{\bibfnamefont{K.-T.} \bibnamefont{Cheng}}, \bibnamefont{and}
  \bibinfo{author}{\bibfnamefont{W.~R.} \bibnamefont{Johnson}},
  \bibinfo{journal}{Phys. Rev. A} \textbf{\bibinfo{volume}{11}},
  \bibinfo{pages}{1946} (\bibinfo{year}{1975}).

\bibitem[{\citenamefont{Volotka et~al.}(2016)\citenamefont{Volotka, Yerokhin,
  Surzhykov, St\"ohlker, and Fritzsche}}]{volotka:2016:023418}
\bibinfo{author}{\bibfnamefont{A.~V.} \bibnamefont{Volotka}},
  \bibinfo{author}{\bibfnamefont{V.~A.} \bibnamefont{Yerokhin}},
  \bibinfo{author}{\bibfnamefont{A.}~\bibnamefont{Surzhykov}},
  \bibinfo{author}{\bibfnamefont{T.}~\bibnamefont{St\"ohlker}},
  \bibnamefont{and}
  \bibinfo{author}{\bibfnamefont{S.}~\bibnamefont{Fritzsche}},
  \bibinfo{journal}{Phys. Rev. A} \textbf{\bibinfo{volume}{93}},
  \bibinfo{pages}{023418} (\bibinfo{year}{2016}).

\bibitem[{\citenamefont{Zhou et~al.}(1992)\citenamefont{Zhou, Kissel, and
  Pratt}}]{zhou:1992:6906}
\bibinfo{author}{\bibfnamefont{B.}~\bibnamefont{Zhou}},
  \bibinfo{author}{\bibfnamefont{L.}~\bibnamefont{Kissel}}, \bibnamefont{and}
  \bibinfo{author}{\bibfnamefont{R.~H.} \bibnamefont{Pratt}},
  \bibinfo{journal}{Phys. Rev. A} \textbf{\bibinfo{volume}{45}},
  \bibinfo{pages}{6906} (\bibinfo{year}{1992}).

\bibitem[{\citenamefont{Fritzsche}(2019)}]{fritzsche:2019:1}
\bibinfo{author}{\bibfnamefont{S.}~\bibnamefont{Fritzsche}},
  \bibinfo{journal}{Comput. Phys. Commun.} \textbf{\bibinfo{volume}{240}},
  \bibinfo{pages}{1} (\bibinfo{year}{2019}).

\bibitem[{\citenamefont{Berestetsky et~al.}(1982)\citenamefont{Berestetsky,
  Lifshitz, and Pitaevsky}}]{berestetsky}
\bibinfo{author}{\bibfnamefont{V.~B.} \bibnamefont{Berestetsky}},
  \bibinfo{author}{\bibfnamefont{E.~M.} \bibnamefont{Lifshitz}},
  \bibnamefont{and} \bibinfo{author}{\bibfnamefont{L.~P.}
  \bibnamefont{Pitaevsky}}, \emph{\bibinfo{title}{Quantum Electrodynamics}}
  (\bibinfo{publisher}{{Pergamon Press, Oxford}}, \bibinfo{year}{1982}).

\bibitem[{\citenamefont{Surzhykov et~al.}(2018)\citenamefont{Surzhykov,
  Yerokhin, Fritzsche, and Volotka}}]{surzhykov:2018:053403}
\bibinfo{author}{\bibfnamefont{A.}~\bibnamefont{Surzhykov}},
  \bibinfo{author}{\bibfnamefont{V.~A.} \bibnamefont{Yerokhin}},
  \bibinfo{author}{\bibfnamefont{S.}~\bibnamefont{Fritzsche}},
  \bibnamefont{and} \bibinfo{author}{\bibfnamefont{A.~V.}
  \bibnamefont{Volotka}}, \bibinfo{journal}{Phys. Rev. A}
  \textbf{\bibinfo{volume}{98}}, \bibinfo{pages}{053403}
  (\bibinfo{year}{2018}).

\bibitem[{\citenamefont{G\"unther et~al.}(2018)\citenamefont{G\"unther,
  Volotka, Jentschel, Fritzsche, St\"ohlker, Thirolf, and
  Zepf}}]{guenther:2018:063843}
\bibinfo{author}{\bibfnamefont{M.~M.} \bibnamefont{G\"unther}},
  \bibinfo{author}{\bibfnamefont{A.~V.} \bibnamefont{Volotka}},
  \bibinfo{author}{\bibfnamefont{M.}~\bibnamefont{Jentschel}},
  \bibinfo{author}{\bibfnamefont{S.}~\bibnamefont{Fritzsche}},
  \bibinfo{author}{\bibfnamefont{T.}~\bibnamefont{St\"ohlker}},
  \bibinfo{author}{\bibfnamefont{P.~G.} \bibnamefont{Thirolf}},
  \bibnamefont{and} \bibinfo{author}{\bibfnamefont{M.}~\bibnamefont{Zepf}},
  \bibinfo{journal}{Phys. Rev. A} \textbf{\bibinfo{volume}{97}},
  \bibinfo{pages}{063843} (\bibinfo{year}{2018}).

\bibitem[{\citenamefont{Samoilenko et~al.}(2020)\citenamefont{Samoilenko,
  Volotka, and Fritzsche}}]{samoilenko:2020:12}
\bibinfo{author}{\bibfnamefont{D.}~\bibnamefont{Samoilenko}},
  \bibinfo{author}{\bibfnamefont{A.~V.} \bibnamefont{Volotka}},
  \bibnamefont{and}
  \bibinfo{author}{\bibfnamefont{S.}~\bibnamefont{Fritzsche}},
  \bibinfo{journal}{Atoms} \textbf{\bibinfo{volume}{8}}, \bibinfo{pages}{12}
  (\bibinfo{year}{2020}).

\bibitem[{\citenamefont{Sapirstein and Johnson}(1996)}]{sapirstein:1996:5213}
\bibinfo{author}{\bibfnamefont{J.}~\bibnamefont{Sapirstein}} \bibnamefont{and}
  \bibinfo{author}{\bibfnamefont{W.~R.} \bibnamefont{Johnson}},
  \bibinfo{journal}{J. Phys. B} \textbf{\bibinfo{volume}{29}},
  \bibinfo{pages}{5213} (\bibinfo{year}{1996}).

\bibitem[{\citenamefont{Shabaev et~al.}(2004)\citenamefont{Shabaev, Tupitsyn,
  Yerokhin, Plunien, and Soff}}]{shabaev:2004:130405}
\bibinfo{author}{\bibfnamefont{V.~M.} \bibnamefont{Shabaev}},
  \bibinfo{author}{\bibfnamefont{I.~I.} \bibnamefont{Tupitsyn}},
  \bibinfo{author}{\bibfnamefont{V.~A.} \bibnamefont{Yerokhin}},
  \bibinfo{author}{\bibfnamefont{G.}~\bibnamefont{Plunien}}, \bibnamefont{and}
  \bibinfo{author}{\bibfnamefont{G.}~\bibnamefont{Soff}},
  \bibinfo{journal}{Phys. Rev. Lett.} \textbf{\bibinfo{volume}{93}},
  \bibinfo{pages}{130405} (\bibinfo{year}{2004}).

\bibitem[{\citenamefont{Kohn and Sham}(1965)}]{kohn:1965:A1133}
\bibinfo{author}{\bibfnamefont{W.}~\bibnamefont{Kohn}} \bibnamefont{and}
  \bibinfo{author}{\bibfnamefont{L.~J.} \bibnamefont{Sham}},
  \bibinfo{journal}{Phys. Rev.} \textbf{\bibinfo{volume}{140}},
  \bibinfo{pages}{A1133} (\bibinfo{year}{1965}).

\bibitem[{\citenamefont{Goldemberg and Pratt}(1966)}]{goldemberg:1966:311}
\bibinfo{author}{\bibfnamefont{J.}~\bibnamefont{Goldemberg}} \bibnamefont{and}
  \bibinfo{author}{\bibfnamefont{R.~H.} \bibnamefont{Pratt}},
  \bibinfo{journal}{Rev. Mod. Phys.} \textbf{\bibinfo{volume}{38}},
  \bibinfo{pages}{311} (\bibinfo{year}{1966}).
  
\bibitem[{\citenamefont{Blum}(2012)}]{blum}
\bibinfo{author}{\bibfnamefont{K.}~\bibnamefont{Blum}},
  \emph{\bibinfo{title}{Density Matrix Theory and Applications}}
  (\bibinfo{publisher}{{Springer, Heidelberg}}, \bibinfo{year}{2012}).

\bibitem[{\citenamefont{Balashov et~al.}(2000)\citenamefont{Balashov,
  Grum-Grzhimailo, and Kabachnik}}]{balashov}
\bibinfo{author}{\bibfnamefont{V.~V.} \bibnamefont{Balashov}},
  \bibinfo{author}{\bibfnamefont{A.~N.} \bibnamefont{Grum-Grzhimailo}},
  \bibnamefont{and} \bibinfo{author}{\bibfnamefont{N.~M.}
  \bibnamefont{Kabachnik}}, \emph{\bibinfo{title}{Polarization and Correlation
  Phenomena in Atomic Collisions}} (\bibinfo{publisher}{{Springer, Boston}},
  \bibinfo{year}{2000}).

\bibitem[{\citenamefont{Tashenov et~al.}(2009)\citenamefont{Tashenov,
  Khaplanov, Cederwall, and Sch{\"a}ssburger}}]{tashenov:2009:599}
\bibinfo{author}{\bibfnamefont{S.}~\bibnamefont{Tashenov}},
  \bibinfo{author}{\bibfnamefont{A.}~\bibnamefont{Khaplanov}},
  \bibinfo{author}{\bibfnamefont{B.}~\bibnamefont{Cederwall}},
  \bibnamefont{and} \bibinfo{author}{\bibfnamefont{K.-U.}
  \bibnamefont{Sch{\"a}ssburger}}, \bibinfo{journal}{Nucl. Instr. and Meth. A}
  \textbf{\bibinfo{volume}{600}}, \bibinfo{pages}{599} (\bibinfo{year}{2009}).

\bibitem[{\citenamefont{Koga and Hayakawa}(2017)}]{koga:2017:204801}
\bibinfo{author}{\bibfnamefont{J.~K.} \bibnamefont{Koga}} \bibnamefont{and}
  \bibinfo{author}{\bibfnamefont{T.}~\bibnamefont{Hayakawa}},
  \bibinfo{journal}{Phys. Rev. Lett.} \textbf{\bibinfo{volume}{118}},
  \bibinfo{pages}{204801} (\bibinfo{year}{2017}).

\end{thebibliography}
\end{document}